\documentclass[%
 reprint,
 superscriptaddress,
 amsmath,amssymb,
 aps,
 floatfix,
]{revtex4-2}
\usepackage{graphicx}
\usepackage{dcolumn}
\usepackage{bm}
\usepackage{hyperref}
\usepackage[mathlines]{lineno}
\usepackage{here}
\usepackage{booktabs}
\usepackage{graphicx} 
\usepackage{siunitx}

\begin{document}


\preprint{APS/123-QED}

\title{Taming Multi-Domain, -Fidelity Data: Towards Foundation Models for Atomistic Scale Simulations}
\author{Tomoya Shiota}
\email{shiota.tomoya.ss@gmail.com}
\affiliation{Graduate School of Engineering Science, The University of Osaka, 1-3 Machikaneyama, Toyonaka, Osaka 560-8531, Japan}
\affiliation{Center for Quantum Information and Quantum Biology, The University of Osaka, 1-2 Machikaneyama, Toyonaka 560-8531, Japan} 
\author{Kenji Ishihara}
\affiliation{Center for Quantum Information and Quantum Biology, The University of Osaka, 1-2 Machikaneyama, Toyonaka 560-8531, Japan} 
\author{Tuan Minh Do}
\affiliation{Center for Quantum Information and Quantum Biology, The University of Osaka, 1-2 Machikaneyama, Toyonaka 560-8531, Japan} 
\author{Toshio Mori}
\affiliation{Center for Quantum Information and Quantum Biology, The University of Osaka, 1-2 Machikaneyama, Toyonaka 560-8531, Japan} 
\author{Wataru Mizukami}
\email{mizukami.wataru.qiqb@osaka-u.ac.jp}
\affiliation{Graduate School of Engineering Science, The University of Osaka, 1-3 Machikaneyama, Toyonaka, Osaka 560-8531, Japan}
\affiliation{Center for Quantum Information and Quantum Biology, The University of Osaka, 1-2 Machikaneyama, Toyonaka 560-8531, Japan} 

\date{\today}

\begin{abstract}

Machine learning interatomic potentials (MLIPs) are changing atomistic simulations in the field of chemistry and materials science. However, constructing a single universal MLIP that can accurately model molecular and crystalline systems remains challenging. A central obstacle is the integration of diverse datasets generated under different computational conditions. We present Total Energy Alignment (TEA), which is an approach that enables the seamless integration of heterogeneous quantum chemical datasets without redundant calculations. 
Using TEA, we trained MACE-Osaka24, the first open-source MLIP model based on a unified dataset covering molecular and crystalline systems.
This universal model displays strong performances across diverse chemical systems, exhibiting similar or improved accuracies in predicting organic reaction barriers compared to those of specialized models, while effectively maintaining state-of-the-art accuracies for inorganic systems. These advancements pave the way for accelerated discoveries in the fields of chemistry and materials science via genuine foundation models for chemistry.

\end{abstract}

\maketitle  


\section{\label{sec:intro}Introduction}

\begin{figure*}[t]
\includegraphics[width=2\columnwidth]{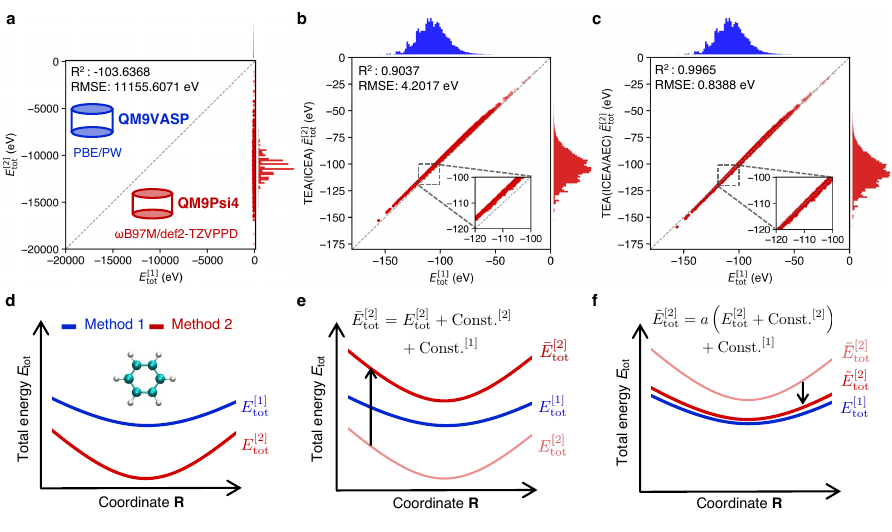}
\caption{\label{fig:1} (a) Scatter plot comparing the total energies of about 143 000 QM9 geometries obtained with Method 1 (PBE/PW via VASP; “QM9VASP”) and Method 2 ($\omega$B97M/def2-TZVPPD via Psi4; “QM9Psi4”). The very poor correlation ($\mathrm{R}^2$ = –103.6, root mean square error (RMSE) = 11 156 eV) underscores the large systematic difference between the two levels of theory; marginal histograms are shown on the axes. (b) Same data after the first stage of Total Energy Alignment (TEA)—Inner Core Energy Alignment (ICEA)—is applied to the total energies of the QM9Psi4. (c) Total energies after the second stage of TEA—Atomization Energy Correction (AEC)—which brings the datasets into close agreement ($\mathrm{R}^2$ = 0.9965, RMSE = 0.839 eV). Insets in (b) and (c) enlarge the boxed regions. (d) Schematic potential-energy surfaces (PESs) for a representative molecule (benzene) calculated with Methods 1 (blue) and 2 (red), corresponding to one data point in (a). (e) Illustration of ICEA: for species with identical stoichiometry, ICEA acts as a constant vertical shift of the Method 2 PES. (f) Illustration of AEC: after ICEA, AEC rescales the shifted Method 2 PES by a factor a, yielding the fully aligned PES that matches the Method 1 reference.}
\end{figure*}

Recent advances in machine learning interatomic potentials (MLIPs) have provided new opportunities in the field of computational chemistry and materials science. Researchers can now perform atomistic simulation with almost first-principles accuracy at computational cost that is orders of magnitude lower~\cite{wang2024machine, kaser2023neural, choudhary2022recent, schmidt2019recent, mishin2021machine, unke2021machine, kocer2022neural, deringer2019machine,chmiela2023accurate, mueller2020machine, liu2023combining,bartok2010gaussian,behler2007generalized,zhang2024exploring,kulichenko2024data}. This paradigm shift has been propelled by increasingly sophisticated architectures—ranging from high-order equivariant neural networks to multiscale graph neural representations—and an expanding wealth of large, first-principles-based datasets~\cite{behler2007generalized,behler2011atom,schutt2018schnet,xie2018crystal,chen2019graph,gasteiger2021gemnet,unke2021spookynet,choudhary2021atomistic,TAKAMOTO2022111280,batzner20223,batatia2022mace,musaelian2023learning,mao2024moleculegraphnetworksmanybody,frank2024euclidean,park2024scalable,cheng2024cartesian,wang2024n,duval2023hitchhiker,jain2013commentary,kirklin2015open,schmidt2022dataset,schmidt2024improving,haastrup2018computational,rosen2021machine,sriram2024open,chanussot2021open,tran2023open,smith2017ani,smith2020ani,schreiner2022transition1x,ramakrishnan2014quantum,nandi2023multixc,nakata2017pubchemqc,nakata2023pubchemqc,eastman2023spice,isert2022qmugs,donchev2021quantum,zhou2019big,ullah2024molecularquantumchemicaldata,gabellini2024openqdc,bochkarev2024graph}. Inorganic-focused MLIPs now span a considerable portion of the periodic table, making it easier to survey crystal structures and discover new phenomena in catalyst, semiconductor, and beyond~\cite{chen2022universal,qi2024robust,takamoto2022towards,deng2023chgnet,choudhary2023unified,batatia2024foundationmodelatomisticmaterials,merchant2023scaling,yang2024mattersimdeeplearningatomistic,barroso2024open,Shiota_GNNTL, shiota2025lowering,neumann2024orb,focassio2024performance,yu2024systematic}. Meanwhile, MLIPs for use in analyzing molecular systems have increased in versatility, realizing almost hybrid density functional theory (DFT) accuracies across a range of organic, pharmaceutical, and biomolecular targets~\cite{smith2017ani1, smith2018less, smith2019approaching,devereux2020extending,zubatyuk2019accurate,anstine2024aimnet2,eastman2023spice,donchev2021quantum,isert2022qmugs,unke2021spookynet,unke2024biomolecular,frank2024euclidean,kabylda2024molecular,kovacs2023mace,batatia2022mace}.

Despite these advances, pursuing a truly universal MLIP that seamlessly unites the organic and inorganic realms remains challenging. Molecular and crystalline datasets often differ in their computational methods, DFT functionals employed, and basis sets, rendering their resulting potential energy surfaces (PESs) incompatible~\cite{schleder2019dft,huang2023central,kresse1993ab,kresse1996efficient,kresse1994ab,kresse1996efficiency,hutter2014cp2k,te2001chemistry,smith2020psi4,ORCA5,ORCA,g16}. For example, inorganic datasets typically use plane-wave basis sets and generalized gradient approximations, whereas organic datasets rely on localized basis sets and hybrid functionals. Merging these heterogeneous sources without recalculating vast portions of the data is challenging, rendering the development of foundation models in chemistry out of reach for numerous researchers with limited computational resources.

Here, we introduce a general strategy denoted Total Energy Alignment (TEA), which addresses this long-standing problem by harmonizing datasets generated using different computational settings.
TEA uses a two-step approach—first aligning the inner-core reference energies and then scaling the atomization energies—to integrate datasets that previously could not be combined. 
Applying TEA to unify large inorganic (MPtrj)~\cite{deng2023chgnet} and broad organic (OFF23~\cite{kovacs2023mace}, consisting of SPICE~\cite{eastman2023spice,donchev2021quantum}, QMug~\cite{isert2022qmugs}, water clusters, and Tripeptides datasets) datasets, we constructed MACE-Osaka24: a single open-source neural network potential that can accurately model organic molecular reactions and extended crystalline systems. Unlike previous multi-task approaches that simply switch between domains~\cite{takamoto2022towards,zhang2024dpa2largeatomicmodel}, 
MACE-Osaka24 addresses organic and inorganic PESs using a single model. 
It not only outperforms specialized potentials in predicting the reaction barriers of drug-like organic molecules but also maintains state-of-the-art accuracies for inorganic systems. 

This study has two key implications. First, by eliminating the necessity of costly recalculations, TEA aids in democratizing the construction of foundation models in chemistry, enabling researchers with limited computational resources to contribute more effectively. Second, MACE-Osaka24 reveals that a single model can exhibit high accuracies across the molecular and inorganic domains, suggesting a new level of interoperability. As data-driven discoveries expands, the capacity to seamlessly handle organic and inorganic chemical spaces is expected to accelerate catalyst design, functional material development, and the exploration of complex reactions. The TEA framework and MACE-Osaka24 point toward truly universal MLIPs, enabling the next generation of foundation models to advance beyond traditional domain boundaries.

\section{\label{sec:results}Results}

\subsection{\label{sec:method:A}Total Energy Alignment (TEA)}

Developing a truly universal MLIP that can handle molecular and extended solid systems requires the unified treatment of datasets generated under diverse computational conditions. However, directly merging these heterogeneous datasets is challenging because their total energies are often incomparable, as shown in Fig.~\ref{fig:1}(a) and (d). Here, we introduce the TEA framework, which is a two-step procedure designed to seamlessly reconcile datasets computed using different quantum chemical approaches, as shown in Fig.~\ref{fig:1}(e) and (f). 

TEA comprises two key steps: (I) Inner Core Energy Alignment (ICEA) and (II) Atomization Energy Correction (AEC). ICEA corrects for systematic energy offsets caused by differences in the treatment of core electrons, such as the use of effective core potentials or projector-augmented wave (PAW)~\cite{blochl1994projector,kresse1999ultrasoft} methods, without altering the relative energy differences. AEC subsequently scales the atomization energies to account for discrepancies in the computational fidelities or basis sets or exchange-correlation functionals used across different datasets. By initially aligning the core-level energies and then applying a scalable correction to the atomization energies, TEA provides a straightforward route for use in integrating previously incompatible datasets into a single coherent training platform, as shown schematically.

In the following explanation of TEA method, we adopt the problem setting and assumptions that a dataset of total energies and forces has been generated with Method 2, labeled [2] in the equations, and that isolated atomic energies for every element in the dataset are available from both Method 1 ([1]) and Method 2. TEA’s goal is to transform the total energy and force dataset produced by Method 2 into the form that would be obtained with Method 1, without recomputing the dataset with Method 1. For simplicity, we now outline the procedure for converting the data of an \(N\)-atom system, taken from the dataset obtained with Method 2, into the corresponding data that would be produced with Method 1, as shown in Fig.~\ref{fig:1}(d)--(f).

\subsubsection{\label{sec:method:A.1}Inner Core Energy Alignment (ICEA)}

Different computational methods often treat inner-core electrons differently, leading to systematic shifts in their total energies. These differences do not generally affect chemical reactivity, but they hamper direct comparisons or combinations of datasets. To address this, we first assume that the relative quantities, such as atomization energies, remain consistent between Methods 1 and 2.

For a system of $N$ atoms, the atomization energy $E_{\mathrm{at}}$ is defined as:
\begin{equation}
E_{\mathrm{at}} = \sum_{i=1}^N E_i^{P_i} - E_{\mathrm{tot}},
\end{equation}
where $E_i^{P_i}$ is the energy of an isolated atom of species $P_i$, and $E_{\mathrm{tot}}$ is the total energy of the system.

Under the assumption that the atomization energies obtained using Methods 1 and 2 are equivalent,
\begin{equation}
E_{\mathrm{at}}^{[1]} = E_{\mathrm{at}}^{[2]},
\end{equation}
the ICEA-shifted total energy of Method 2, $\bar{E}_{\mathrm{tot}}^{[2]}$, is given as
\begin{align}
\bar{E}_{\mathrm{tot}}^{[2]}
&= \sum_{i=1}^N E_i^{P_i,[1]} - E_{\mathrm{at}}^{[2]} \nonumber \\
&= \left(\sum_{i=1}^N E_i^{P_i,[1]} - \sum_{i=1}^N E_i^{P_i,[2]}\right) + E_{\mathrm{tot}}^{[2]} .
\end{align}
This relation shows that we can shift the total energies from Method 2 onto the reference scale of Method 1 using only isolated-atom energies. In practice, ICEA sets a common baseline for both datasets, ensuring that differences arise from meaningful chemical effects rather than arbitrary computational choices.

\subsubsection{\label{sec:method:A.2}Atomization Energy Correction (AEC)}

After applying ICEA, certain residual differences in the atomization energies still remain if the two datasets originate from different calculation protocols (e.g., distinct levels of theory, different basis sets, or contrasting exchange-correlation functionals). These differences manifest as systematic offsets that must be corrected before the datasets can be fully integrated.

We introduce a correction function $f$ that relates the atomization energies obtained via the two methods:
\begin{equation}
E_{\mathrm{at}}^{[1]} = f\left(E_{\mathrm{at}}^{[2]}\right).
\end{equation}
To maintain simplicity and ensure a robust performance, we adopt a single scaling factor $a$:
\begin{equation}
f\left(E_{\mathrm{at}}^{[2]}\right) = a \, E_{\mathrm{at}}^{[2]}.
\end{equation}
In fact, previous studies have reported a linear relationship between the magnitude of the atomization energy and the systematic errors present~\cite{savin2015judging,perdew2016intensive}, making a simple scaling approach a practical choice. Using the correction function defined by the scaling factor \(a\), the AEC-aligned total energy $\tilde{E}_{\mathrm{tot}}^{[2]}$ is expressed as follows:
\begin{equation}
\tilde{E}_{\mathrm{tot}}^{[2]} = \sum_{i=1}^N E_i^{P_i,[1]} - a \, E_{\mathrm{at}}^{[2]}.
\end{equation}
Because forces \(\{\mathbf{F}_i\}_{i=1}^N\) are negative gradients of the total energy \(E_\mathrm{tot}\) with respect to atomic coordinates \(\{\mathbf{R}_i\}_{i=1}^N\) , this correction consistently adjusts forces $\mathbf{F}_i^{[2]}$ as well:
\begin{equation}
\mathbf{\tilde{F}}_i^{[2]} = -\frac{\partial \tilde{E}_{\mathrm{tot}}^{[2]}}{\partial \mathbf{R}_i} = -a \frac{\partial E_{\mathrm{tot}}^{[2]}}{\partial \mathbf{R}_i} = a\mathbf{F}_i^{[2]},
\end{equation}
where $\mathbf{\tilde{F}}_i^{[2]}$ is AEC-aligned forces. This ensures that the entire PES is appropriately rescaled. Together, ICEA and AEC yield a coherent PES alignment that preserves relative energy differences and accuracies across heterogeneous datasets.

\subsubsection{\label{sec:method:A.3}TEA for QM9 datasets}

To evaluate the performance of TEA between datasets that employed the different fidelity functionals and differed in core electron treatments, basis sets, and boundary conditions, we conducted TEA for QM9 datasets re-calculated using the VASP~\cite{kresse1993ab,kresse1996efficient,kresse1994ab,kresse1996efficiency} and Psi4~\cite{smith2020psi4}, named QM9VASP and QM9Psi4, respectively.

Fig.~\ref{fig:1}(a) shows a parity plot comparing QM9VASP and QM9Psi4. No clear trend in the total energy between the two datasets is observed, and the data points are scattered. This is because QM9Psi4 is calculated using an all-electron method, where the total energy is the energy of all electrons, whereas QM9VASP represents the total energy of the valence electrons only. As shown in Fig.~\ref{fig:1}(b), by performing TEA using ICEA, we succeed in aligning the total energies to be comparable. However, the precision is as high as the root mean square error (RMSE) 4.2017~eV and the reliability is low. This is mainly because of the differences in fidelity caused by the different functionals. As shown in Fig.~\ref{fig:1}(c), TEA using ICEA/AEC captures the systematic differences due to fidelity variations and significantly improves the RMSE to 0.8388~eV.


\subsection{Evaluation of Universal MLIPs}

\begin{figure*}[]
\includegraphics[]{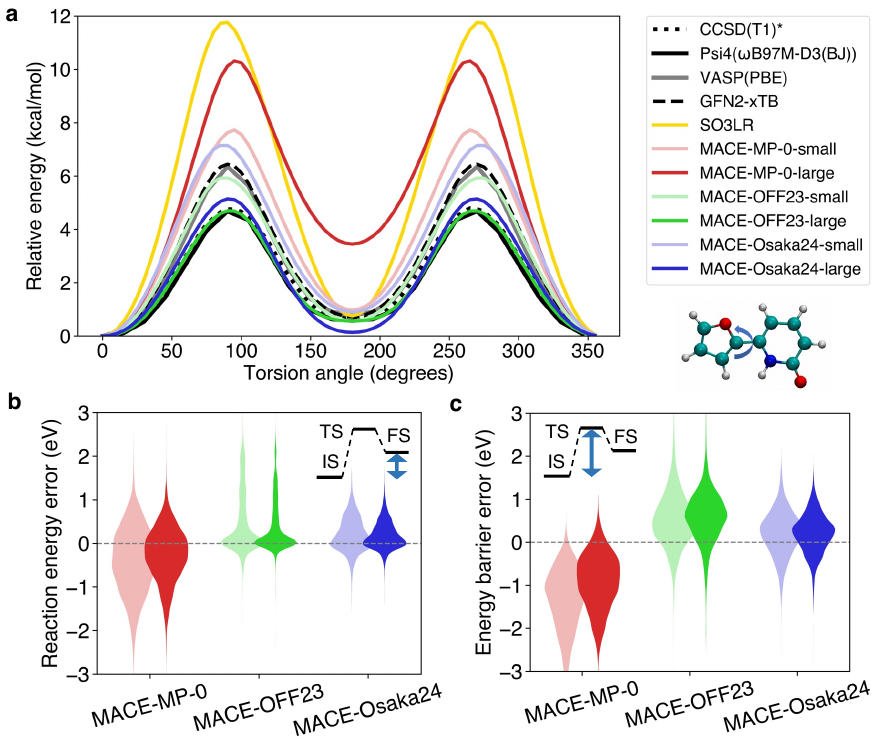}
\caption{\label{fig:2} (a) Optimized torsional potential energy surfaces (PESs) of dihedral torsion in a representative organic molecule of the biaryl torsion dataset~\cite{lahey2020benchmarking} shown on the right-hand side the of the figure. The results obtained using various Machine Learning Interatomic Potentials (MLIPs), including the SO3LR, MACE-MP-0, MACE-OFF23, and MACE-Osaka24 models, are compared alongside reference calculations performed using Psi4 ($\mathrm{\omega}$B97M-D3(BJ)), the VASP (PBE), and ORCA  (CCSD(T1)*). The CCSD(T1)* values were obtained from the biaryl torsion benchmark~\cite{lahey2020benchmarking}. (b) Violin plot of the errors in the reaction energies, where the reaction energy is defined as the energy difference between the initial (IS) and final states (FS). The errors are calculated based on single-point energy calculations of the 10 073 organic reactions of the Transition1x dataset conducted using the MACE-MP-0, MACE-OFF23, and MACE-Osaka24 models. The results were compared to the single-point energies calculated at the $\omega$B97M-D3(BJ) level using Psi4 . The results obtained using the large and small models are respectively shown in darker and lighter colors. (c) Violin plot of the errors in the energy barriers, where the energy barrier is defined as the energy difference between the IS and TS. The results were compared to the single-point energies at the $\omega$B97M-D3(BJ) level of the 10 073 organic reactions of the Transition1x dataset, as calculated using Psi4 and the same models as those shown in (b). The lighter and darker colors represent the results obtained using the small and large models, respectively .}
\end{figure*}

\begin{figure*}[]
\includegraphics[]{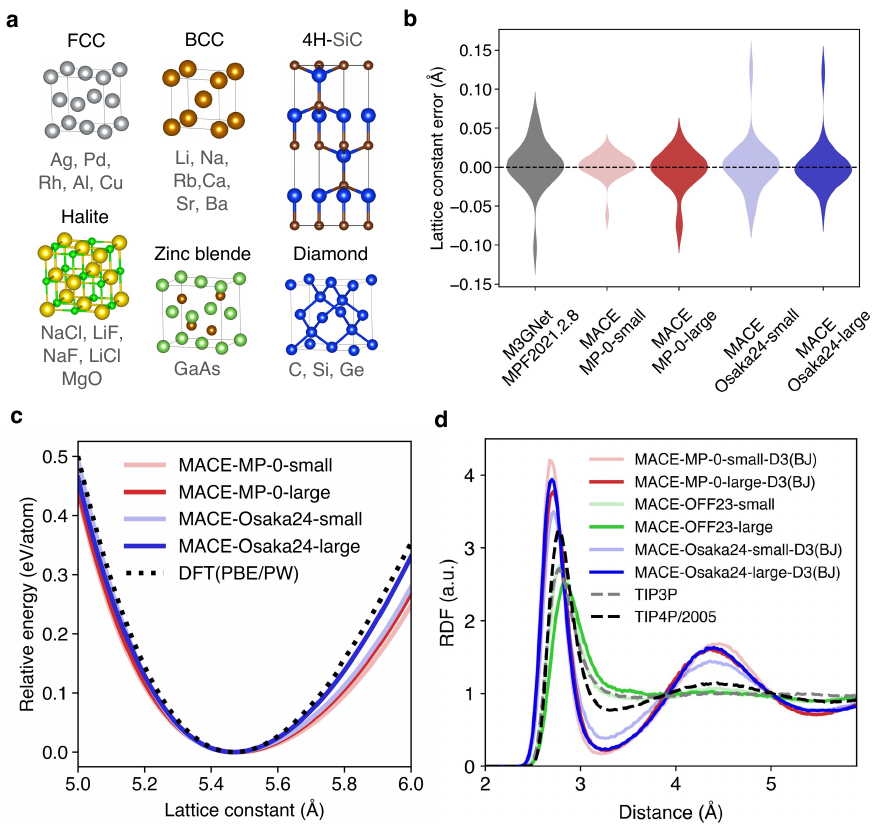}
\caption{\label{fig:3}(a) Crystal structures and their representative materials used in the lattice constant benchmark shown in (b): Face-centered cubic (FCC, e.g., Ag, Pd), body-centered cubic (BCC, e.g., Li, Na), halite (e.g., NaCl), zinc blende (e.g., GaAs), and Diamond (e.g., C, Si). (b) Violin plot showing the errors in the lattice constants ($\mathrm{\AA}$) obtained using different models, including MACE-MP-0-small, MACE-MP-0-large, MACE-Osaka24-small, MACE-Osaka24-large, and M3GNet trained on the MPF.2021.2.8 dataset. The errors are calculated with respect to lattice constants optimized using the VASP with the PBE functional, employing the MPRelaxSet input provided by Pymatgen from the Materials Project. (c) Relative energy (eV/atom) of a diamond-structured Si crystal as a function of the lattice constant ($\mathrm{\AA}$), as predicted using MACE models (MP-0 and Osaka24 variants) and compared to that predicted via VASP calculations. The VASP calculations were performed using the MPStaticSet input provided by Pymatgen. (d) Radial distribution function (RDF, a.u.) of liquid water obtained via NVT simulations. The results obtained using the MACE-MP-0 and MACE-Osaka24 models with D3(BJ) corrections are shown, in addition to those obtained via classical MD simulations using the TIP3P and TIP4P/2005 water models.}
\end{figure*}

\begin{figure*}[]
\includegraphics[]{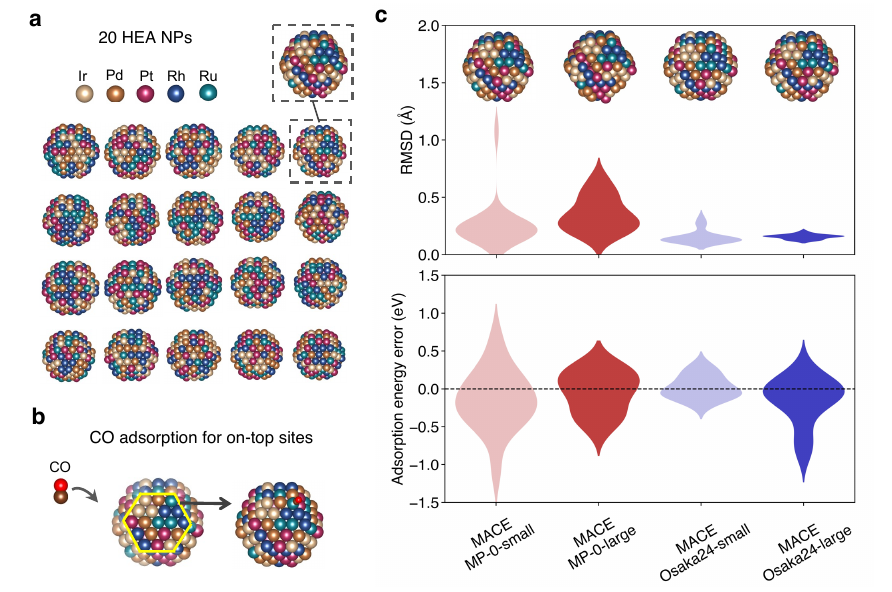}
\caption{\label{fig:8}(a) Structures of twenty equiatomic IrPdPtRhRu high-entropy alloy nanoparticles (HEA NPs) with 201 atoms obtained from PBE-level DFT geometry optimizations with VASP.
(b) Schematic of CO adsorption at on-top sites of a HEA NP surface. The yellow hexagon highlights a target CO molecular adsorption sites. CO adsorption-energies were obtained from Ref.~\cite{shiota2025lowering} for 17 of the 19 on-top sites that PBE-level DFT calculations identified as stable. The example shows a CO molecule adsorbed in an on-top configuration on a Ru corner atom. (c) Benchmarking of the NP systems with both the small and large variants of the MACE-MP-0 and MACE-Osaka24 models. The upper violin plot shows the distribution of root mean square deviations (RMSDs) for the twenty equiatomic IrPdPtRhRu HEA NP structures in (a), while the lower plot presents the error distribution of CO on-top adsorption energies relative to PBE-level DFT calculations provided in Ref.~\cite{shiota2025lowering}.}
\end{figure*}

We evaluated the performances of the constructed multi-domain universal MLIPs, i. e., MACE-Osaka24-small and MACE-Osaka24-large. For comparison, we also performed simulations using MACE-MP-0 and MACE-OFF23, where viable. Furthermore, we present the results of simulation obtained using other universal MLIPs, in addition to DFT and semi-empirical and classical force fields as additional references. Appendix~\ref{sec:appendix:F} presents a validation of the training results using the M3GNet model to demonstrate that our approach generalizes beyond the MACE architecture.

\begin{table}[b]
\centering
\caption{Mean absolute errors (MAEs) of the barrier heights of 78 drug-like biaryl torsions compared to the high-fidelity reference energies. The values between parentheses were obtained from a previous study~\cite{kovacs2023mace}.}
\label{tab:1}
\begin{tabular}{lc}
\hline
Universal MLIP        & MAE (kcal/mol) \\ \hline
MACE-OFF23-large       & 0.403 (0.3)          \\
MACE-Osaka24-large       & 0.457          \\
MACE-OFF23-small       & 0.598 (0.5)         \\
MACE-Osaka24-small       & 0.695          \\
GFN2-xTB               & 0.898          \\
MACE-MP-0-large        & 1.909          \\
SO3LR                  & 2.451          \\
MACE-MP-0-small        & 3.386          \\ \hline
\end{tabular}
\end{table}

\begin{table}[h!]
\centering
\caption{Mean absolute errors (MAEs) of the predicted reaction energies and energy barriers of 10 073 reactions in the Transition1x dataset. The units are all in eV.}
\label{tab:2}
\begin{tabular}{lcc}
\hline
Universal MLIPs & Reaction energy & Energy barrier \\
\hline
MACE-Osaka24-large & 0.265 & 0.404   \\
MACE-Osaka24-small & 0.336 & 0.457   \\
MACE-OFF23-large & 0.436 & 0.711   \\
MACE-OFF23-small & 0.544 & 0.672   \\
MACE-MP-0-large  & 0.519 & 0.937   \\
MACE-MP-0-small  & 0.686 & 1.333   \\
\hline
\end{tabular}
\label{tab:mae_mean_values}
\end{table}

\subsubsection{Molecular Systems}

First, we present the benchmark results for organic molecular systems. Table~\ref{tab:1} shows the mean absolute errors (MAEs) of the barrier heights of 78 drug-like biaryl torsions compared to high-fidelity reference energies at the coupled cluster level of theory provided in the biaryl torsion benchmark~\cite{lahey2020benchmarking}. Compared to the root mean square errors (RMSEs) in the benchmark reported by Kovács et al.~\cite{kovacs2023mace}, the difference of approximately 0.1 kcal/mol in the accuracies of the MACE-OFF23 models is likely attributable to differences in the optimizer used for the torsional PES calculations. The details of the biaryl torsion benchmark, including the computational settings for DFT and the universal MLIPs, are described in Appendix~\ref{sec:appendix:C.3}. The MACE-Osaka24-small and -large models yield  predictions that are 2.69 and 1.45~kcal/mol more accurate, respectively, in the molecular torsion reactions compared to those of MACE-MP-0-small and -large. As revealed in the study by Kovács et al.~\cite{kovacs2023mace}, the MACE-OFF23 models and semi-empirical GFN2-xTB~\cite{bannwarth2019gfn2} method provide quantitative predictions with chemical accuracies (1 kcal/mol) in terms of calculations at the coupled cluster level of theory. Similarly, the MACE-Osaka24 models yield chemical accuracies, indicating their effectiveness in precisely predicting molecular torsion. Fig.~\ref{fig:2}(a) shows the torsional PESs of a molecule in the biaryl torsion benchmark. The MACE-MP-0-large model overestimates the barrier height of the torsion reaction by approximately double. Compared to the PES calculated at the PBE level using the VASP, the difference is large. This suggests that quantitatively predicting organic molecular domains is challenging using MLIPs trained only with inorganic crystal domains. The MACE-OFF23-small model provides predictions that are almost equivalent to the results at the PBE level. The MACE-Osaka24-small model displays a predictive performance almost equivalent to that of MACE-MP-0-small. The MACE-Osaka24-large model exhibits a predictive accuracy close to that of the high-fidelity $\omega$B97M-D3(BJ) model, exceeding the predictive accuracy at the PBE level. These results suggest that the integration of learning datasets using TEA enables a single model to inherit the accuracies of the original datasets.

\begin{table}[b]
\caption{Mean absolute errors (MAEs) of the lattice constants predicted using the universal machine learning interatomic potentials (MLIPs) compared to those predicted via PBE-level DFT calculations for bulk crystals.}
\label{tab:3}
\begin{tabular}{lcc}
\hline
Universal MLIP & MAE (\AA) \\
\hline
MACE-MP-0-small & 0.0100 \\
MACE-Osaka24-large & 0.0148 \\
MACE-Osaka24-small & 0.0166 \\
MACE-MP-0-large & 0.0166 \\
\hline
\end{tabular}
\label{tab:mae_comparison}
\end{table}

\begin{table}[b]
\caption{Average root mean square deviations (RMSDs) of the geometries predicted using the MLIPs compared to those predicted via PBE-level DFT calculations for 20 quinary IrPdPtRhRu HEA NPs.}
\label{tab:4}
\begin{tabular}{lcc}
\hline
Universal MLIP & RMSD (\AA) \\
\hline
MACE-Osaka24-small & 0.154 \\
MACE-Osaka24-large & 0.156 \\
MACE-MP-0-small & 0.259 \\
MACE-MP-0-large & 0.356 \\

\hline
\end{tabular}
\end{table}

\begin{table}[b]
\caption{Root mean square errors (RMSEs) of the adsorption energies predicted using the universal MLIPs compared to those predicted via PBE-level DFT calculations for CO on-top adsorption on HEA NPs.}
\label{tab:5}
\begin{tabular}{lcc}
\hline
Universal MLIP & RMSE (eV) \\
\hline
MACE-Osaka24-small & 0.152 \\
MACE-MP-0-large & 0.274 \\
MACE-Osaka24-large & 0.341 \\
MACE-MP-0-small & 0.414 \\

\hline
\end{tabular}
\end{table}

We then evaluated the performances of the universal MLIPs using the Transition1x dataset, focusing on their capacities to predict the reaction energies and energy barriers of 10 073 organic reactions. The details of the calculations are provided in Appendix~\ref{sec:appendix:C.4}. Fig.~\ref{fig:2}(b) shows the distributions of the prediction errors in the reaction energies, where MACE-Osaka24 exhibits the lowest error spread compared to those of MACE-MP-0 and MACE-OFF23. Similarly, Fig.~\ref{fig:2}(c) shows the prediction errors in the energy barriers, indicating that MACE-Osaka24 consistently outperforms the other models, particularly in capturing transition state (TS) regions with higher accuracies. Table~\ref{tab:2} quantitatively supports this observation. The MAEs of the reaction energies predicted using MACE-MP-0-small, MACE-OFF23-small, and MACE-Osaka24-small are 0.686, 0.544, and 0.336 eV, respectively, and the respective MAEs of the predicted energy barriers are 1.333, 0.672, and 0.457 eV. The larger models of each potential exhibit further improvements, with MACE-Osaka24-large yielding the lowest MAEs of 0.265 and 0.404 eV for the reaction energies and energy barriers, respectively. Hence, MACE-Osaka24, particularly its large model, offers superior predictive accuracies in predicting the reaction energies and energy barriers of the Transition1x dataset. This highlights the importance of tailored model architectures and training datasets that explicitly include TS regions, enabling MLIPs to realize high accuracies, even for reactive systems far from equilibrium.

\subsubsection{Condensed-Phase Systems}

We then verified the accuracies of the universal MLIPs for the crystalline systems shown in Fig.~\ref{fig:3}(a). The crystals used for the benchmark were those adopted in Section B.4 of the Supporting Information in the study by Batatia et al. ~\cite{batatia2022mace}. The details of the calculation conditions for the crystal benchmarks are provided in Appendix~\ref{sec:appendix:C.5}. The benchmark results for each crystal and crystal structure are analyzed in detail in Appendix~\ref{sec:appendix:D}. Fig.~\ref{fig:3}(b) shows the error distributions of the predictions of the lattice constants calculated using various universal MLIPs and the VASP at the same computational level as that of the training data, i.e., the PBE functional. The MAEs of the MACE-Osaka24 models exceed those of the MACE-MP-0 models. However, as shown in Table~\ref{tab:3}, the MAEs of the predictions of the MACE-MP-0 and MACE-Osaka24 models are lower than those of the predictions of the pretrained M3GNet model. The differences in predictive accuracy between MACE-MP-0 and MACE-Osaka24 are 0.008~\AA\ and 0.002~\AA\ for the small and large models, respectively. This suggests that integrating the data for organic molecules with different fidelities and domains using TEA does not reduce the original predictive accuracy. Fig.~\ref{fig:3}(c) shows the PES of the lattice constant of diamond-structured Si. The models accurately predict the equilibrium lattice constant at the PBE level calculated using the VASP (calculation conditions of the MPStaticSet of the Materials Project). Furthermore, the PES description is superior using MACE-Osaka24 compared to that using MACE-MP-0 with respect to the results of VASP calculation. This is likely coincidental but indicates the high levels of robustness of multidomain universal MLIPs.

We then evaluated the performance of the universal MLIPs for liquid water at room temperature (\SI{300}{K}). Fig.~\ref{fig:3}(d) shows the radial distribution function (RDF) of the O--O atoms obtained via the MD simulation of bulk liquid water at room temperature (\SI{300}{K}), which is critical for organic and inorganic materials. The MACE-MP-0 and MACE-Osaka24 models apply the D3(BJ) correction. Details of the MD calculations using the MLIPs and classical force fields are provided in Appendix~\ref{sec:appendix:C.6}. The MACE-OFF23 model describes the properties of liquid water at room temperature~\cite{kovacs2023mace}. MACE-MP-0-D3(BJ) reproduces the RDF at the PBE-D3(BJ) level~\cite{batatia2024foundationmodelatomisticmaterials}. MACE-Osaka24-large-D3(BJ) provides RDF descriptions that are almost equivalent to those of MACE-MP0-D3(BJ). In contrast, MACE-Osaka24-small-D3(BJ) provides an RDF that is approximately intermediate between those of MACE-MP-0 and MACE-OFF23. This suggests that the capacity to describe dynamic properties changes significantly, depending on the complexity of the architecture and balance of the dataset. 

\subsubsection{Nanoparticle Catalyst Systems}

Finally, we evaluated the generalization capability of the universal MLIPs on transition-metal nanoparticle (NP) catalysts that were absent from the training set. The test cases comprised twenty 201-atom, truncated octahedral IrPdPtRhRu high-entropy alloy (HEA) NPs~\cite{Wu_2020} and CO adsorption at on-top sites on a representative HEA NP as shown in Fig.~\ref{fig:8}(a) and (b), respectively. The PBE-level DFT calculation results were provided in Ref.~\cite{shiota2025lowering}.  Computational details are provided in the Appendix. The average root mean square deviations (RMSDs) across twenty HEA NPs in Table~\ref{tab:4} show that the MACE-Osaka24 models reproduce NP structures more accurately than the MACE-MP-0 models, a trend corroborated by the narrower RMSD distributions in the upper violin plot of Fig.~\ref{fig:8}(c). For adsorption energies (Table~\ref{tab:5} and the lower panel of Fig.~\ref{fig:8}(c)), MACE-Osaka24-small achieved the best accuracy, with an RMSE of 0.152 eV—improving upon MACE-MP-0-small by 0.26 eV. In contrast, the RMSE of Osaka24-large was 0.07 eV higher than that of MP-0-large. Overall, MACE-Osaka24-small most closely reproduces PBE-level DFT results for both NP geometries and CO adsorption energies. These results suggest that training on TEA-aligned, combined molecular and crystalline datasets enhances predictive performance for out-of-domain systems such as heterogeneous catalytic surfaces.

\section{\label{sec:discussion}Discussion}

The results indicate that TEA is an effective method of combining different datasets. By aligning the inner-core reference energies and adjusting the atomization energies, TEA bridges the differences caused by differing computational details, such as basis sets and exchange-correlation functionals. Using TEA, we merged the MPtrj inorganic crystal dataset with the OFF23 organic dataset to train the MACE-Osaka24—a multi-domain MLIP that exhibits an accuracy comparable to those of specialized models, such as MACE-MP-0 and MACE-OFF23, while covering a considerably wider range of chemical systems.

The key advantage of TEA is that it simplifies data integration without changing the architecture of the MLIP. Unlike methods such as $\Delta$-machine learning~\cite{ramakrishnan2015big} or multi-fidelity learning~\cite{chen2021learning, kim2024dataefficientmultifidelitytraininghighfidelity, liu2024high}, which often target specific domains or fidelities, TEA offers a general, straightforward method of combining datasets. This approach enables the use of existing data from various sources without extensive recalculations. By demonstrating that a single model—MACE-Osaka24—can accurately predict the energies of molecular reactions, lattice constants of inorganic crystals, and properties of liquid water, we confirmed that the resulting PESs maintained their physical consistencies and meaningful energy gradients across diverse chemical environments.

However, certain limitations and challenges remain unresolved. The current implementation relies on the availability of suitable reference atomic energies and geometries, which can be more challenging for systems with strong electron correlations, charged species, or relativistic effects. While using a single global scaling factor for the atomization energies was successful in this study, certain specialized cases may require more nuanced correction schemes. 

\section{\label{sec:conclusion}Conclusion}

We introduce the TEA methodology as a robust, efficient framework for use in unifying heterogeneous quantum chemical datasets into a single-level PES. Using TEA, we constructed a single universal MLIP, i.e., MACE-Osaka24, which exhibited state-of-the-art accuracies for molecular and crystalline systems. Its performance was comparable to those of specialized models, such as MACE-MP-0 for inorganic solids and MACE-OFF23 for organic molecules, without costly recalculations under a single theoretical framework.

However, the effect of TEA extends beyond its technical contributions. The integration of diverse datasets without high-cost recalculations aids in democratizing the development of foundation models in the field of chemistry. This approach aligns with the shift toward open science, wherein the use of a wide range of data sources is increasingly essential. As the chemistry and materials science communities continue to produce larger, more varied datasets, TEA provides a practical route towards truly universal MLIPs, accelerating the identification of materials, drugs, and catalysts via collaborative, data-driven research.

Future research may evaluate TEA using datasets obtained via higher-level quantum chemical methods or directly include correlation and relativistic effects. Continued advances in neural network architectures, training methods, and hyperparameter optimization are likely to improve the levels of robustness and accuracies of universal MLIPs. As research communities produce larger, more varied first-principles datasets, the concepts demonstrated by TEA and MACE-Osaka24 can guide the development of more fully integrated and widely accessible foundation models. Such models, which are firmly based on reliable first-principles accuracy yet are adaptable to different computational approaches, are expected to aid us in exploring and understanding increasingly complex chemical systems.

\section{\label{sec:methods}Methods}

\subsection{\label{sec:method:B}Datasets}

To demonstrate the effectiveness of TEA, we integrated two large-scale datasets: the MPtrj dataset, which provides inorganic structures calculated at the PBE~\cite{perdew1996generalized} functional with plane-wave basis sets (PBE/PW) using the Vienna Ab initio Simulation Package (VASP)~\cite{kresse1993ab,kresse1996efficient,kresse1994ab,kresse1996efficiency}, and the OFF23 dataset, which is an extensive organic dataset computed at the $\omega$B97M-D3(BJ)/def2-TZVPPD~\cite{najibi2018nonlocal,grimme2010consistent,grimme2011effect,weigend2005balanced} level using Psi4~\cite{smith2020psi4}. Prior to integration, we removed the D3(BJ) dispersion correction from the OFF23 data to avoid double-counting the dispersion effects in the final MLIP.

To determine the scaling factor $a$ used in the AEC and assess the uncertainties, we also employed the QM9 dataset~\cite{ramakrishnan2014quantum}, originally computed at B3LYP~\cite{stephens1994ab}/6-31G(2df,p) level using Gaussian09~\cite{g16}. We recalculated QM9 using the VASP (PBE/PW) and Psi4 ($\omega$B97M-D3(BJ)/def2-TZVPPD) to generate QM9VASP and QM9Psi4 subsets, ensuring consistent reference points for establishing $a$. The full details of dataset preparation and integration, including the corrections and final merged sets, are provided in Appendix~\ref{sec:appendix:A}. The fully integrated organic-inorganic dataset is publicly available at \url{https://github.com/qiqb-osaka/mace-osaka24}.

\subsection{\label{sec:method:C}MLIP Training}

Using TEA-enabled integration, we trained the MLIPs using the MACE framework~\cite{batatia2022mace,batatia2024foundationmodelatomisticmaterials}, employing mace v0.3.6 (\url{https://github.com/ACEsuit/mace}). We leveraged the integrated MPtrj/OFF23 dataset after applying TEA, and we denote the resulting MLIP as MACE-Osaka24. The model and the final training data are available at \url{https://github.com/qiqb-osaka/mace-osaka24}.

Training followed the hyperparameters, cost functions, and optimizers of the MACE-MP-0-small and MACE-MP-0-large models described in Ref.~\cite{batatia2024foundationmodelatomisticmaterials}, with modifications. For all models, we set a cutoff radius of 4.5~\AA{} for constructing the atomic neighborhood graph. We used isolated atomic energies computed with spin polarization utilizing the VASP as references for the atomic species within OFF23. Model training was performed using 32 A100 graphics processing units (GPUs) in parallel. Details of the MACE-Osaka24 model training procedure and the hyperparameter settings can be found in Appendix~\ref{sec:appendix:B}.

\begin{acknowledgments}
This project was supported by funding from the MEXT Quantum Leap Flagship Program (MEXTQLEAP) through Grant No. JPMXS0120319794, the JST COI-NEXT Program through Grant No. JPMJPF2014, and the JST ASPIRE Program Grant No. JPMJAP2319. The completion of this research was partially facilitated by the JSPS Grants-in-Aid for Scientific Research (KAKENHI), specifically Grant Nos. JP23H03819 and JP21K18933. We thank the Supercomputer Center, the Institute for Solid State Physics, the University of Tokyo, for allowing us to use their facilities. A part of the calculations were performed using the Genkai supercomputer of the Research Institute for Information Technology at Kyushu University. This work was also achieved using the SQUID supercomputer at the Cybermedia Center, The University of Osaka.
\end{acknowledgments}

\begin{verbatim}
\end{verbatim}


\nocite{*}

\bibliography{apssamp}

\appendix

\section{\label{sec:related}Related Work}

The goal of an MLIP is to realize first-principles accuracy in simulating chemical and materials systems while significantly reducing computational costs. Early methods, such as Behler-Parrinello networks and Gaussian approximation potentials, revealed that machine learning could reproduce high-level quantum chemistry results without directly solving the Schrödinger equation for each geometry~\cite{behler2007generalized,bartok2010gaussian}. More advanced E(3)-equivariant graph neural networks and message-passing models have since emerged, improving both accuracy and transferability~\cite{schutt2018schnet,xie2018crystal,gasteiger2021gemnet,unke2021spookynet,choudhary2021atomistic,batzner20223,batatia2022mace,musaelian2023learning}. Concurrently, large-scale first-principles datasets—ranging from the extensive inorganic databases of the Materials Project~\cite{jain2013commentary,kirklin2015open,chen2022universal,deng2023chgnet,batatia2024foundationmodelatomisticmaterials} to molecular sets, such as the QM9~\cite{ramakrishnan2014quantum}, OFF23~\cite{eastman2023spice,isert2022qmugs,donchev2021quantum,kovacs2023mace}, and SPICE~\cite{eastman2023spice} datasets -- have enabled the training of increasingly universal MLIPs. Consequently, models such as MACE-MP-0~\cite{batatia2022mace,batatia2024foundationmodelatomisticmaterials} and CHGNet~\cite{deng2023chgnet} currently display state-of-the-art performances for inorganic crystals, whereas others, such as MACE-OFF23~\cite{kovacs2023mace} and AIMNet2~\cite{zubatyuk2019accurate,anstine2024aimnet2}, deliver high accuracies across diverse organic and biomolecular systems.

A major challenge in advancing universal MLIPs is the integration of heterogeneous datasets constructed using different computational protocols, basis sets, and exchange-correlation functionals into a single cohesive training set. These differences affect the reference energies, force field definitions, and inclusion of periodic conditions within the calculations, rendering the direct combination of the data challenging.~\cite{schleder2019dft,huang2023central} To date, several strategies have been applied in attempting to bridge these discrepancies, e.g., $\Delta$-machine learning and multi-fidelity learning approaches learn corrections from lower- to higher-level references, allowing them to blend datasets at different accuracies. \cite{ramakrishnan2015big,chen2021learning,liu2024high,kim2024dataefficientmultifidelitytraininghighfidelity} 
However, these methods often require a reference dataset covering the fidelity and domain ranges and still face difficulties when data are generated using different software or fundamentally different computational setups. Consequently, numerous solutions remain specialized for specific domains, e.g., molecular systems or periodic solids, but not both simultaneously.

Only a handful of attempts exist to span the organic and inorganic domains using a single MLIP have been reported, e.g., PFP~\cite{takamoto2022towards} uses multi-task learning to handle molecular and crystalline data together, but it treats them as separate 'modes' rather than unifying their energy scales. Similarly, DPA-2~\cite{zhang2024dpa2largeatomicmodel} improves generalization via pretraining on multiple tasks—including molecules, crystals, and surfaces—but it still depends on carefully managed workflows and fine-tuning rather than the direct merging of heterogeneous datasets. These approaches highlight the advantages of multi-domain learning, such as improved transferability, fewer data requirements, and stronger PES exploration. However, they are yet to resolve the core problem of integrating data generated under different computational conditions into one consistent PES without extensive recalculations.

Another avenue of research seeks to to align different datasets using physically meaningful reference values. For inorganic materials, methods like the Fitted Elemental Reference Energies approach compare the formation and elemental reference energies of inorganic materials across various exchange-correlation functionals and calculation setups~\cite{jain2011formation,stevanovic2012correcting,wang2021framework,kingsbury2022flexible} Recently, Gabellini et al.\cite{gabellini2024openqdc} introduced a large molecular dataset by converting total energies to atomization energies (analogous to formation energies), which aids in reducing the reliance on absolute reference values that differ between computational codes. However, the atomization energies exhibit systematic errors, depending on the computational protocol~\cite{savin2015judging,perdew2016intensive}. Consequently, simply transforming the current datasets into atomization energies does not guarantee more effective MLIP training. Although these strategies offer promising leads, applying them in integrating large-scale organic and inorganic datasets, where the computational fidelities and natures of the systems (extended solids versus finite molecules) differ, remains non-trivial.

\begin{figure*}[]
\includegraphics[]{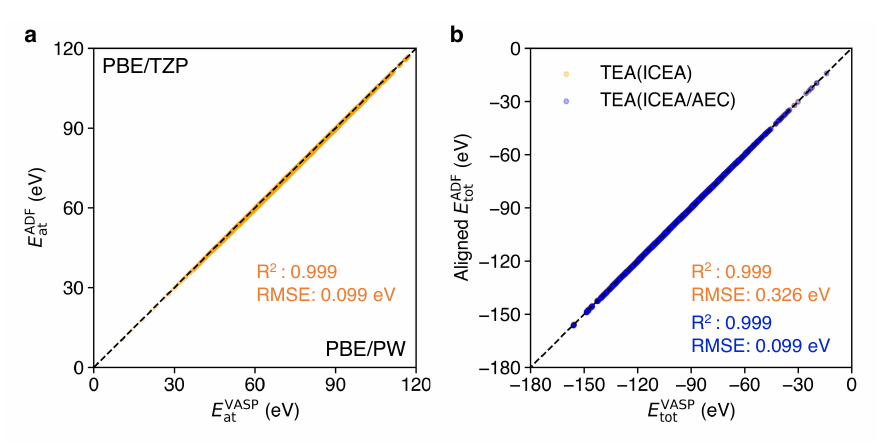}
\caption{\label{fig:4}Results of the total energy alignment (TEA) of different datasets. (a) Parity plot of the atomization energies of the QM9VASP and QM9ADF datasets, as calculated using the same PBE functional. (b) Parity plot of the total energies after applying Inner Core Energy Alignment (ICEA) and Atomization Energy Correction (AEC) to the QM9ADF dataset.}
\end{figure*}

\begin{figure*}[]
\includegraphics[]{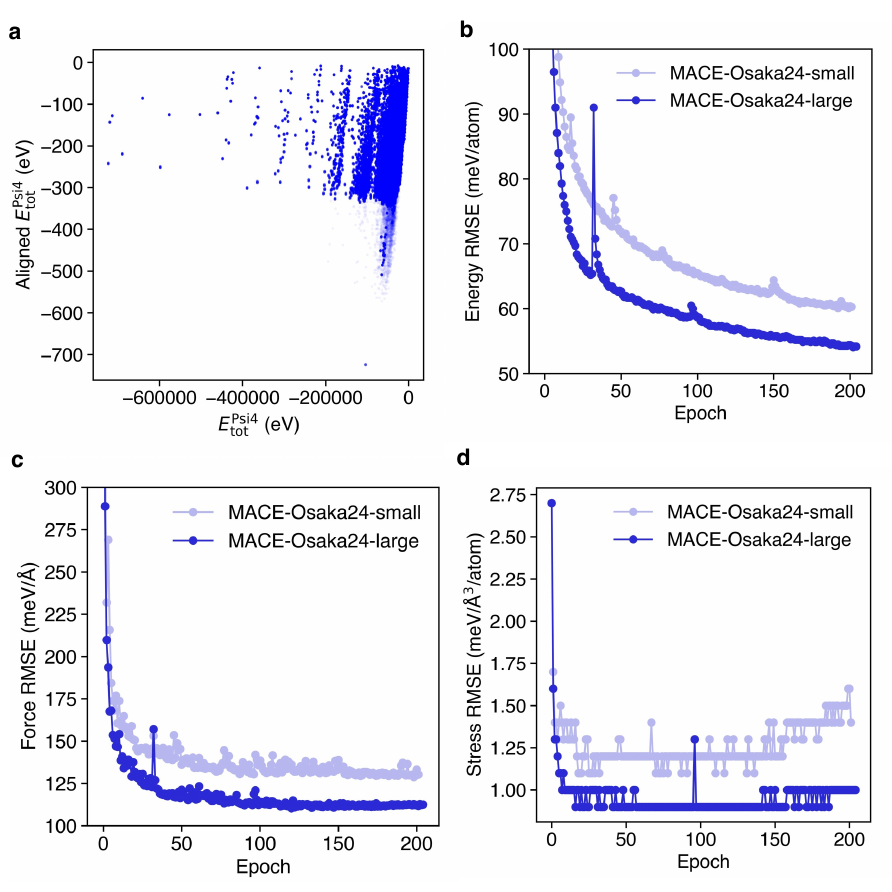}
\caption{\label{fig:5}(a) Parity plot of the total energies obtained using the original OFF23 dataset and those obtained after applying total energy alignment (TEA). (b) Root mean square errors (RMSEs) of the energies during the training of the MACE-Osaka24-small and MACE-Osaka24-large models over ~200 epochs. (c) RMSEs of the forces during the training of the same models. (d) RMSEs of the stresses during the training of the same models.}
\end{figure*}

\section{\label{sec:appendix:A}Total Energy Alignment (TEA) of the QM9 Dataset}


To evaluate the performance of TEA between datasets that employed the same fidelity functionals and differed in core electron treatments, basis sets, and periodic boundary conditions, we conducted TEA using the VASP~\cite{kresse1993ab,kresse1996efficient,kresse1994ab,kresse1996efficiency} and Amsterdam Density Functional (ADF)~\cite{te2001chemistry}. We utilized the QM9 dataset, which comprises approximately 134 000 molecules optimized at the B3LYP/6-31G(d) level using Gaussian09~\cite{ramakrishnan2014quantum}. By performing single-point energy calculations at the Perdew-Burke-Ernzerhof/plane-wave basis set (PBE/PW) level with VASP using the Gaussian09-optimized geometries, we generated a new dataset denoted QM9VASP. For the TEA target, we adopted the PBE/TZP level dataset from MultiXC-QM9~\cite{nandi2023multixc}, as calculated using ADF with various functionals, which excluded the molecules involved in charge separation~\cite{kim2019energy}. Hereafter, we refer to this dataset as QM9ADF.

Although QM9VASP and QM9ADF were computed with the same exchange–correlation functional, differences in basis sets and core-electron treatment lead to a root mean square error (RMSE) of 0.6156 eV between the two datasets. By contrast, the comparison of atomization energies between QM9VASP and QM9ADF, illustrated in Fig.~\ref{fig:4}(a), yields an RMSE nearly an order of magnitude lower, at just 0.0992 eV. As illustrated in Fig.~\ref{fig:4}(b), applying TEA with Inner Core Energy Alignment (ICEA) aligns the total energies of QM9VASP and QM9ADF, roughly halving the discrepancy and reducing the RMSE to 0.3258 eV. Adding Atomization Energy Correction (AEC) on top of ICEA (i.e., ICEA/AEC) lowers the total-energy RMSE even further, to 0.0992 eV—matching the error observed for the atomization energies.

\section{\label{sec:appendix:B}Training Multi-domain Universal MLIPs}

We demonstrated that the stable training of universal MLIPs was possible by integrating datasets of the organic domain, to which TEA was applied, into datasets of the inorganic domain. In this study, as shown in Fig. ~\ref{fig:5}(a), we constructed a TEA-integrated MPtrj/OFF23 dataset, which used the AEC scaling factor using QM9Psi4 and QM9VASP. For the MLIP architecture, we adopt the small and large MACE architectures proposed by Batatia et al.~\cite{batatia2024foundationmodelatomisticmaterials}. The constructed multidomain universal MLIPs are denoted MACE-Osaka24-small and MACE-Osaka24-large. Fig.~\ref{fig:5}(b)--(d) show the learning curves of the MACE-Osaka24 models for the energy, force, and stress. The RMSEs of the large model, with a larger size, are generally smaller than those of the small model. This is highly consistent with the trends of the learning curves reported in the study by Batatia et al.~\cite{batatia2024foundationmodelatomisticmaterials}.

\section{\label{sec:appendix:C}Computational Details}

\subsection{\label{sec:appendix:C.1}QM9VASP Dataset Generation}

The QM9VASP dataset was generated using the VASP version 5.4.4. For the exchange-correlation functional, we adopted the PBE functional to generate the MPtrj dataset. A plane-wave energy cutoff of 400 eV was employed to expand  the electronic wavefunctions (ENCUT = 400). The geometries were obtained from the original QM9 dataset. To prevent interactions between adjacent unit cells, a vacuum layer of 10 \text{\AA} was introduced into each unit cell. The electronic self-consistency loop converged when the total energy change between successive iterations was less than \(1 \times 10^{-5}\) eV (EDIFF = 1e-05). The symmetry operations were disabled (ISYM = 0). High-precision settings were used throughout the calculations (PREC = accurate) to ensure reliable results. The Brillouin zone integrations were performed using the Gaussian smearing method with a smearing width of 0.1 eV (ISMEAR = 0, SIGMA = 0.1). 

\subsection{\label{sec:appendix:C.2}QM9Psi4 Dataset Generation}

The QM9Psi4 dataset was generated using Psi4 version 1.9. To ensure that the computational conditions were equivalent to those employed in generating the OFF23 dataset, we adopted the $\omega$B97M-D3(BJ) functional, which adds the D3 dispersion correction with the Becke-Johnson (BJ) damping function to the exchange-correlation functional $\omega$B97M. The def2-TZVPPD basis set was used in the calculations. When determining the scaling factor for the AEC step of TEA, we omitted the D3(BJ) dispersion correction.

\subsection{\label{sec:appendix:C.3}Biaryl Torsion Benchmark}
First, we introduced the method of generating a biaryl torsion dataset reported by Lahey et al.~\cite{lahey2020benchmarking}, which provides reference energies at the coupled-cluster level, as shown in Table~\ref{tab:1}. The torsional potential energy surfaces (PESs) were computed using second-order density-fitting Møller-Plesset perturbation theory (DF-MP2) with the def2-TZVP basis set (DF-MP2/def2-TZVP). The CCSD(T1)*/CBS energies were obtained by combining DLPNO-CCSD(T) (denoted CCSD(T)*)~\cite{guo2018communication}, the complete basis set (CBS) correction scheme proposed by Smith et al.\cite{smith2019approaching,lahey2020benchmarking}, and iterative triple CCSD(T1) methods~\cite{guo2018communication}. These torsional PES values served as reference data in this study.

Torsional PES optimizations of the dihedral torsions of the 78 molecules presented in the biaryl torsion benchmark~\cite{lahey2020benchmarking} were performed using GFN2-xTB~\cite{bannwarth2019gfn2}, MACE-MP-0, SO3LR~\cite{kabylda2024molecular}, MACE-OFF23, MACE-Osaka24, the VASP, and Psi4. In the calculations using GFN2-xTB and the MLIPs, the dihedral angles were varied in increments of 5°, and structural relaxations were performed with each dihedral angle constrained at its set value. Using the VASP and Psi4, the increment of the dihedral angle was set to 15°. In all methods, geometry relaxations were conducted until the force acting on each atom was less than 0.01 eV/\AA. The optimization calculations of the constrained geometries were conducted using Atomic Simulation Environment version 3.23.0~\cite{larsen2017atomic}. Kovács et al.~\cite{kovacs2023mace} performed torsional PES optimization using the TorsionDrive~\cite{qiu2020driving} algorithm.

\subsection{\label{sec:appendix:C.4}Benchmark on Transition1x}

The original Transition1x dataset was generated using ORCA version 5.0.2 with the exchange functional $\omega$B97x and the basis set 6–31G(d). Since our constructed MACE-Osaka24 models are based on the OFF23 dataset computed with Psi4, we performed single-point calculations on the initial state, transition state, and final state geometries of the 10 073 reactions provided in Transition1x dataset using Psi4 under the computational conditions specified in Appendix~\ref{sec:appendix:C.2} to ensure compatibility. Similarly, all validations using MLIPs and GFN2-xTB were carried out by performing single-point calculations on the IS, TS, and FS geometries provided in Transition1x.

\begin{table*}[htbp]
    \centering
    \caption{MAEs of lattice constants (in \AA) predicted utilizing the universal MLIPs compared to those predicted utilizing DFT with the PBE functional using the VASP, as categorized by the type of crystal structure. The predictions of the M3GNet-MPF2021.2.8, MACE-MP-0-small, MACE-MP-0-large, MACE-Osaka24-small, and MACE-Osaka24-large models are evaluated across various structures, including 4H, BCC, diamond, FCC, halite, and zinc blende structures. The errors correspond to the deviations shown in Fig.~\ref{fig:6}(b).}
    \label{tab:6}
    \begin{tabular}{lccccc}
        \toprule
        {Crystal structure} & {M3GNet-MPF2021.2.8} & {MACE-MP-0-small} & {MACE-MP-0-large} & {MACE-Osaka24-small} & {MACE-Osaka24-large} \\
        \midrule
        4H & 0.0143 & 0.0036 & 0.0032 & \bf{0.0002} & 0.0033 \\
        BCC & \bf{0.0701} & 0.3453 & 0.2153 & 0.1536 & 0.1423 \\
        Diamond & 0.0100 & 0.0047 & 0.0031 & \bf{0.0020} & 0.0029 \\
        FCC & 0.0200 & \bf{0.0096} & 0.0179 & 0.0118 & 0.0156 \\
        Halite & 0.0111 & \bf{0.0036} & 0.0108 & 0.0113 & 0.0100 \\
        Zinc blende & \bf{0.0017} & 0.0153 & 0.0205 & 0.0049 & 0.0031 \\
        \hline
    \end{tabular}
    \label{tab:mae_per_structure}
\end{table*}

\subsection{\label{sec:appendix:C.5}Bulk Crystal Lattice Constant}

\begin{figure*}[]
\includegraphics[scale=1.2]{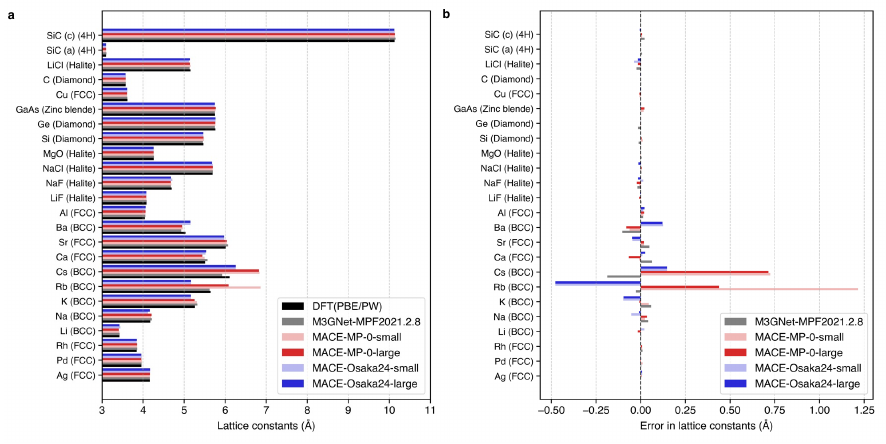}
\caption{\label{fig:6}(a) Lattice constants of bulk crystals, as predicted utilizing the universal MLIPs and DFT calculations using the PBE functional with the VASP. The theoretical level is the same as those of the training data of the universal MLIPs presented in Section~\ref{sec:results}. (b) Prediction errors of the universal MLIPs relative to the lattice constants predicted using DFT.}
\end{figure*}

The crystals used for the benchmark were those adopted in Section B.4 of the Supporting Information in the study by Batatia et al. ~\cite{batatia2024foundationmodelatomisticmaterials}. Representing the body-centered cubic (BCC) materials K, Rb, and Cs using the cutoff radius of 4.5 Å employed in graph construction for the MACE-Osaka24 models was impossible, and thus, they were excluded for the benchmark in Section~\ref{sec:results}. Further details are provided in Appendix~\ref{sec:appendix:D}. The lattice constants were optimized via first-principles calculations using the VASP at the PBE level. The input parameters from MPRelaxSet in the Pymatgen~\cite{ong2013python} library of the Materials Project were utilized to ensure compatibility with the MPtrj dataset used to train the MACE-MP-0 and MACE-Osaka24 models. For the MACE-MP-0 and MACE-Osaka24 models, the convergence criterion for unit cell optimization was set to 0.01 eV/{\AA}. First-principles PES calculations for Si as functions of the lattice constant, as shown in Fig.~\ref{fig:3}(c), were performed using single-point calculations with inputs from the MPStaticSet in Pymatgen. The lattice constants were varied from 5 {\AA} to 6 {\AA} in increments of 0.01 {\AA}.

\subsection{\label{sec:appendix:C.6}Molecular Dynamics (MD) of Liquid Water}

For machine learning (ML)-driven MD simulations, the interface with MACE was implemented using a modified version of OpenMM-ML~\cite{eastman2023openmm}, enabling the incorporation of D3(BJ) dispersion force corrections. For ML-driven MD, a PBC box containing 64 H\(_2\)O molecules at a density of 1 g/cm\(^3\) and NVT ensemble simulations were performed for 100 ps, with the final 50 ps used in the analysis. The equations of motions were integrated using Langevin dynamics with the LFMiddle discretization with a temperature of 298K, a friction coefficient of 1 ps$^{-1}$, and a time step of 1 fs~\cite{zhang2019unified}. 

Classical MD simulations were performed using GROMACS version 2023.3~\cite{abraham2015gromacs} with the TIP3P~\cite{jorgensen1983comparison} and TIP4P/2005~\cite{abascal2005general} water models. The TIP4P/2005 model was used as a reference in evaluating the accuracies of the MLIPs because it reproduces the thermodynamic properties of water with high accuracies over a wide temperature range~\cite{valle2024accuracy}. The leap-frog algorithm with a time step of 2 fs was used. For classical MD, simulations were performed in an NVT ensemble with 1,000 H\(_2\)O molecules at a density of 1 g/cm\(^3\) for 1,000 ps, with the final 500 ps used in the analysis. The temperature was set to 298 K by coupling the system to a v-rescale thermostat with a coupling time of 0.1 ps~\cite{bussi2007canonical}. The smooth particle–mesh Ewald summation was employed to compute the electrostatic interactions with a real-space cutoff of 1.4 nm, an interpolation order of 4, a relative tolerance of $10^{-5}$, and a Fourier spacing of 0.12 nm~\cite{essmann1995smooth}. For the Lennard-Jones potential, the cutoff was set to 1.4 nm, a constant shift was applied such that the potential was zero at the cutoff, and long-range dispersion corrections for energy and pressure were included. The SETTLE algorithm was adopted to treat the water molecules as rigid~\cite{miyamoto1992settle}.

\subsection{\label{sec:appendix:C.7}High-Entropy Alloy Nanoparticle Catalysts}

Twenty equiatomic IrPdPtRhRu HEA NPs,
each containing 201 atoms in a truncated-octahedral shape, were taken from the Figshare repository 
\url{https://doi.org/10.6084/m9.figshare.26973409.v2} that accompanies Fig.~4 of Ref.~\cite{shiota2025lowering}. All first-principles reference data were generated with
\textsc{VASP} 5.4.4, employing the PBE for exchange--correlation functional and the PAW method. Electronic states were converged with the blocked Davidson algorithm in a spin-restricted framework,
a plane-wave cutoff of 400 eV, and a $\Gamma$-point $k$-mesh. Atomic positions were relaxed with a conjugate-gradient scheme until the maximum force fell below $0.01\;\text{eV}/\,\text{\AA}$. For CO adsorption, we considered the 17 on-top sites that Ref.~\cite{shiota2025lowering} identified as PBE-stable, out of the 19 symmetry-distinct atop positions on each NP. Adsorption energies, $E_{\mathrm{ad}}$s, were evaluated as
\begin{equation}
E_{\mathrm{ad}}
= E(\mathrm{CO/NP})
- E(\mathrm{CO})
- E(\mathrm{NP}),
\end{equation}
where $E(\mathrm{CO/NP})$ is the total energy of the relaxed adsorbate–nanoparticle complex,
$E(\mathrm{CO})$ the energy of an isolated CO molecule, and $E(\mathrm{NP})$ the energy of the bare nanoparticle. Universal MLIPs were benchmarked by re-optimizing each bare and CO-covered NP with \textsc{ASE}. The CO molecule was initially positioned 2\,\AA\ above the target surface atom along the resultant vector obtained by summing the vectors defined as the coordinate differences between each nearest-neighbor atom and the adsorption-site atom. A constrained BFGS relaxation
(force threshold $=0.01\;\text{eV}/\,\text{\AA}$)
was then performed, allowing only the adsorbate and the adsorption-site atom to move along the resultant vector,
while all other degrees of freedom were fully relaxed.

\section{\label{sec:appendix:D}Crystal Lattice Constants}

\begin{figure}[]
\centering
\includegraphics[scale=1.1]{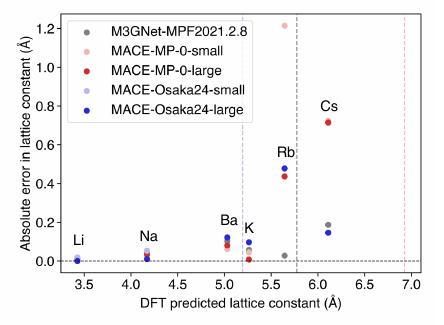}
\caption{\label{fig:7}Absolute errors in the predicted lattice constants of BCC-type crystals according to various universal MLIPs as functions of the DFT-predicted lattice constants. The models compared are M3GNet-MPF2021.2.8, MACE-MP-0 (small and large), and MACE-Osaka24 (small and large). The cutoff radius used for graph construction in the MACE-MP-0 models are 6.0 Å, and its permissible range of lattice constants is shown by the red dashed line. For the M3GNet-MPF2021.2.8 model, the cutoff radius is 5.0 Å, with the corresponding lattice-constant range indicated by the gray dashed line. The MACE-Osaka24 models employ a 4.5 Å cutoff radius, and its lattice-constant range is marked by the blue dashed line. Specific elements (Li, Na, K, Rb, Cs, Ba) are labeled for clarity.}
\end{figure}

\begin{figure*}[]
\centering
\includegraphics{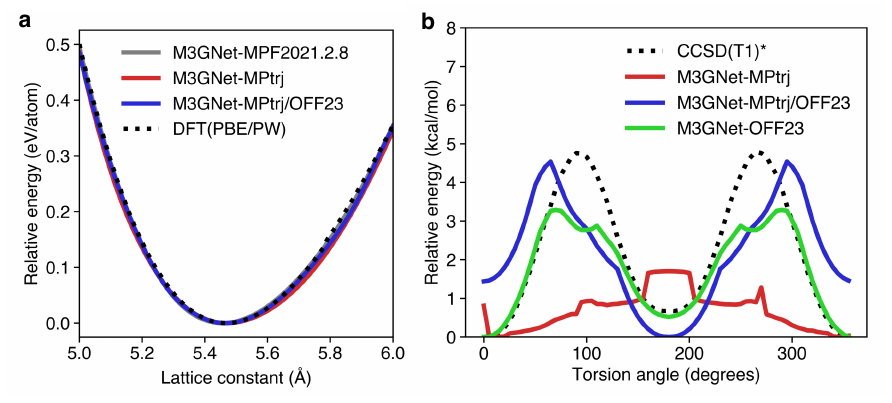}
\caption{\label{fig:9}(a) Relative energy (eV/atom) of a diamond-structured Si crystal as a function of the lattice constant ($\mathrm{\AA}$), as predicted using M3GNet models and compared to that predicted via VASP calculations. The VASP calculations were performed using the MPStaticSet input provided by Pymatgen. (b) Optimized torsional potential energy surfaces (PESs) of dihedral torsion in a representative organic molecule of the biaryl torsion dataset~\cite{lahey2020benchmarking} shown on the right-hand side the of the figure. The results obtained using M3GNet models, are compared alongside reference calculations performed using ORCA at CCSD(T1)* level. The CCSD(T1)* values were obtained from the biaryl torsion benchmark~\cite{lahey2020benchmarking}.}
\end{figure*}

In this section, we analyze the results of the predictions of the lattice constants of representative bulk crystals presented in the Results, in addition to the predictive performance for each crystal structure. Fig.~\ref{fig:6} shows the lattice constants predicted using density functional theory (DFT) and the universal MLIPs. As shown in Fig.~\ref{fig:6}(b), the predictions of M3GNet-MPF2021.2.8 exhibit the fewest outliers among those of the predictions of the models. As shown in Table~\ref{tab:6}, the M3GNet-MPF2021.2.8, MACE-MP-0, and MACE-Osaka24 MLIPs display prediction accuracies with mean absolute errors (MAEs) of less than 0.02 \(\text{Å}\) for crystal structures other than BCC structures. However, the prediction errors for the BCC-type crystal structures exceed those for the other crystal structures. The poor predictive performances for BCC crystals may be related to the cutoff radius used in graph construction. Fig.~\ref{fig:7} shows the DFT-predicted lattice constants of BCC crystals and absolute errors in the predicted lattice constants. As the lattice constant increases, the prediction errors of the models also increase. MACE-Osaka24 was built with a cutoff radius of 4.5 Å, which limits the BCC lattice constants it can capture to a maximum of 5.196 Å (blue dashed line).
M3GNet-MPF 2021.2.8 and MACE-MP-0 were constructed with cutoff radii of 5.0 Å and 6.0 Å , giving upper bounds of 5.774 Å (gray dashed line) and 6.928 Å (red dashed line), respectively. When lattice constants exceed these limits, isolated-atom configurations or simple-cubic structures are superimposed, and the BCC framework is no longer properly represented. Hence, MACE-Osaka24 cannot be used for predicting the lattice constants of K, Rb, and Cs, while M3GNet-MPF 2021.2.8 is unsuitable only for Cs. Although the cutoff radii of M3GNet-MPF 2021.2.8 and MACE-MP-0 are larger than the target BCC lattice constants, their predicted values generally converge toward their respective cutoff limits, indicating that even larger cutoff radii may be required for further accuracy improvements.

\section{\label{sec:appendix:F}Validation of Effectiveness of the TEA Protocol Using M3GNet}

\begin{table}[b]
\caption{Mean absolute errors (MAEs) of the lattice constants predicted using the M3GNet machine learning interatomic potentials (MLIPs) compared to those predicted via PBE-level DFT calculations for bulk crystals.}
\label{tab:7}
\begin{tabular}{lcc}
\hline
Universal MLIP & MAE (\AA) \\
\hline
M3GNet-MPtrj & 0.0099 \\
M3GNet-MPF2021.2.8 & 0.0192 \\
M3GNet-MPtrj/OFF23 & 0.0213 \\
\hline
\end{tabular}
\end{table}

\begin{table}[h!]
\centering
\caption{Mean absolute errors (MAEs) of the predicted reaction energies and energy barriers of 10 073 reactions in the Transition1x dataset using the M3GNet machine learning interatomic potentials (MLIPs). The units are all in eV.}
\label{tab:8}
\begin{tabular}{lcc}
\hline
Universal MLIPs & Reaction energy & Energy barrier \\
\hline
M3GNet-MPtrj/OFF23  & 0.487 & 0.769   \\
M3GNet-OFF23  & 0.979 & 1.315   \\
M3GNet-MPtrj  & 0.766 & 1.625   \\
\hline
\end{tabular}
\end{table}

To demonstrate that the TEA protocol is architecture-agnostic, we trained several M3GNet models with matgl v1.2.7~\cite{ko2025materialsgraphlibrarymatgl}. Using the OFF23, MPtrj, and TEA-integrated MPtrj/SPICE datasets, we obtained three models—denoted M3GNet-OFF23, M3GNet-MPtrj, and M3GNet-MPtrj/OFF23—while keeping all hyper-parameters consistent with those reported for the publicly released M3GNet-MP-2021.2.8-PES model. Training employed a Huber loss in which energy, force, and stress contributions were weighted 1 : 1 : 0.1. Each model was optimized with Adam for 50 epochs using a batch size of 32. Reference atomic energies were taken as the isolated atomic energies calculated with the same settings as each training dataset. All models were trained on the Genkai supercomputer using eight NVIDIA H100 graphics processing units (GPUs) per run. 

Figure~\ref{fig:9}(a) presents the PES for diamond Si crystal. The M3GNet-MPF2021.2.8, the newly trained M3GNet-MPtrj, and the M3GNet-MPtrj/OFF23 via TEA protocol all reproduce accurately the PBE-level DFT reference. In the accompanying lattice constant benchmark shown in Table~\ref{tab:7}, M3GNet-MPtrj lowers the MAE by roughly 0.01 Å relative to M3GNet-MPF2021.2.8. The M3GNet-MPtrj/OFF23 attains accuracy comparable to other M3GNet and MACE models summarized in Tables~\ref{tab:2} and~\ref{tab:7}.

Figure~\ref{fig:9}(b) shifts the focus to torsional PESs for the molecule introduced in Figure~\ref{fig:2}(a). Whereas MACE yields a smoothly varying curve, the M3GNet-MPtrj exhibits a jagged landscape and fails to capture the qualitative dihedral trend. On the other hand, both M3GNet-OFF23 and, more notably, the M3GNet-MPtrj/OFF23 recover a smooth torsional profile that tracks the reference with reasonable accuracy.

The benefits of TEA extend to molecular reactions. Table~\ref{tab:8} compiles MAEs for reaction energies and reaction barriers across the 10 073 systems in Transition1x. The M3GNet-MPtrj/OFF23 surpasses its inorganic-specific M3GNet-MPtrj and organic-specific M3GNet-OFF23 counterparts, mirroring the improvements observed for MACE models.

Taken together, these results demonstrate that TEA—a protocol that reconciles energy scales from heterogeneous data—enables improving performance across crystalline and molecular systems, without relying on a specific neural-network architecture. The recent debut of Allegro-FM introduced by K. Nomura et al., trained on the TEA-integrated MPtrj/OFF23 dataset, provides further confirmation of the architecture-agnostic nature of TEA~\cite{nomura2025allegrofmequivariantfoundationmodel}. As foundation models broaden their scope to encompass the full spectrum of materials and molecular chemistry, TEA is poised to remain the key energy-alignment strategy that underpins their accuracy and transferability.
\end{document}